\documentclass[preprint]{aastex}
\usepackage{amssymb}
\usepackage{amsmath}
\usepackage{graphicx}

\newcommand{\txd}{{\text{d}}}

\newcommand{\bfk}{{\boldsymbol{k}}}
\newcommand{\bfr}{{\boldsymbol{r}}}

\begin{document}

\title{Radiative equilibrium in Monte Carlo radiative transfer using frequency distribution adjustment}

%\author{Maarten~Baes}
%\affil{Sterrenkundig Observatorium, Universiteit Gent, Krijgslaan
%281-S9, B-9000 Gent, Belgium}
%\affil{European Southern Observatory, Casilla 19001, Santiago, Chile}
%\email{maarten.baes@ugent.be}
%
%\author{Dimitris~Stamatellos}
%\affil{Department of Physics and Astronomy, Cardiff University,
%PO Box 913, Cardiff CF24 3YB, UK}
%\email{dimitrios.stamatellos@astro.cf.ac.uk}
%
%\author{Jonathan~I.~Davies}
%\affil{Department of Physics and Astronomy, Cardiff University,
%PO Box 913, Cardiff CF24 3YB, UK}
%\email{jonathan.davies@astro.cf.ac.uk}
%
%\author{Anthony~P.~Whitworth}
%\affil{Department of Physics and Astronomy, Cardiff University,
%PO Box 913, Cardiff CF24 3YB, UK}
%\email{anthony.whitworth@astro.cf.ac.uk}
%
%\author{Sabina~Sabatini}
%\affil{Osservatorio Astronomico di Roma, Via di Frascati 33,
%00040 Monte Porzio Catone, Italy}
%\email{sabatini@mporzio.astro.it}
%
%\author{Sarah~Roberts}
%\affil{Department of Physics and Astronomy, Cardiff University,
%PO Box 913, Cardiff CF24 3YB, UK}
%\email{sarah.roberts@astro.cf.ac.uk}
%
%\author{Suzanne~M.~Linder} 
%\affil{Space
%Telescope Science Institute, 3700 San Martin Drive, Baltimore, MD
%21218, USA}
%\email{linder@stsci.edu} 
%\and
%\author{Rhodri~Evans}
%\affil{Centre for Astronomy and Space Education, University of
%Glamorgan, 4 Forest Grove, Treforest, Pontypridd CF37 1DL, UK}
%\email{rhevans@glam.ac.uk}

\author{%
Maarten~Baes\altaffilmark{1,2}, 
Dimitris Stamatellos\altaffilmark{3},
Jonathan I.~Davies\altaffilmark{3},
Anthony~P.~Whitworth\altaffilmark{3},
Sabina~Sabatini\altaffilmark{4},
Sarah~Roberts\altaffilmark{3},
Suzanne~M.~Linder\altaffilmark{5}, 
Rhodri~Evans\altaffilmark{6}
} 

\altaffiltext{1}{Sterrenkundig Observatorium, Universiteit Gent, Krijgslaan
281-S9, B-9000 Gent, Belgium}
\altaffiltext{2}{European Southern Observatory, Casilla 19001, Santiago 19, Chile, , {\tt{mbaes@eso.org}}}
\altaffiltext{3}{Department of Physics and Astronomy, Cardiff University,
PO Box 913, Cardiff CF24 3YB, UK}
\altaffiltext{4}{Osservatorio Astronomico di Roma, Via di Frascati 33,
00040 Monte Porzio Catone, Italy}
\altaffiltext{5}{Space
Telescope Science Institute, 3700 San Martin Drive, Baltimore, MD
21218, USA}
\altaffiltext{6}{Centre for Astronomy and Space Education, University of
Glamorgan, 4 Forest Grove, Treforest, Pontypridd CF37 1DL, UK}

\begin{abstract}
The Monte Carlo method is a powerful tool for performing radiative
equilibrium calculations, even in complex geometries. The main
drawback of the standard Monte Carlo radiative equilibrium methods is
that they require iteration, which makes them numerically very
demanding. Bjorkman \& Wood recently proposed a frequency distribution
adjustment scheme, which allows radiative equilibrium Monte Carlo
calculations to be performed without iteration, by choosing the
frequency of each re-emitted photon such that it corrects for the
incorrect spectrum of the previously re-emitted photons. Although the
method appears to yield correct results, we argue that its theoretical
basis is not completely transparent, and that it is not completely
clear whether this technique is an exact rigorous method, or whether
it is just a good and convenient approximation. We critically study
the general problem of how an already sampled distribution can be
adjusted to a new distribution by adding data points sampled from an
adjustment distribution. We show that this adjustment is not always
possible, and that it depends on the shape of the original and desired
distributions, as well as on the relative number of data points that
can be added. Applying this theorem to radiative equilibrium Monte
Carlo calculations, we provide a firm theoretical basis for the
frequency distribution adjustment method of Bjorkman \& Wood, and we
demonstrate that this method provides the correct frequency
distribution through the additional requirement of radiative
equilibrium. We discuss the advantages and limitations of this
approach, and show that it can easily be combined with the presence of
additional heating sources and the concept of photon
weighting. However, the method may fail if small dust grains are
included, or if the absorption rate is estimated from the mean
intensity of the radiation field.
\end{abstract}

\keywords{radiative transfer --
radiation mechanisms: thermal --
dust: extinction --
methods: numerical
}

\section{Introduction}

The Monte Carlo method is a very powerful method to solve complicated
radiative transfer problems. Apart from allowing virtually any
geometrical distribution of sources and sinks, it has the potential to
address a number of additional problems which form a serious challenge
to the more conventional ray-tracing techniques. Possible problems
that can be addressed include the calculation of polarization of
scattered radiation (Code \& Whitney~1995; Bianchi et al.~1996), the
correct treatment of kinematical information in the radiative transfer
problem (Mathews \& Wood~2001; Baes \& Dejonghe~2002; Baes et
al.~2003), and inclusion of the clumpiness of the interstellar medium
(Boiss\'e~1990; Witt \& Gordon~1996; Bianchi et al.~2000b).

In this paper we concentrate on another typical radiative transfer
problem, the self-consistent heating and re-emission of an
absorbing/scattering medium in thermal equilibrium with the radiation
field. When dust grains\footnote{In principle, any heating and opacity
source can be accounted for, as long as the optical properties of the
opacity source are independent of temperature. We will focus on the
heating of dust grains which are heated by an ambient stellar
radiation field.} absorb (mainly optical or UV) radiation from the
ambient radiation field, they are heated, and they re-emit this
absorbed energy at longer wavelengths, thereby altering the ambient
radiation field.  Hence the radiative transfer problem requires a
simultaneous calculation of both the temperature distribution of the
dust and the ambient radiation field.

As long as the geometry of the system is not too complicated, this
problem can be solved with conventional techniques. In spherical
geometry, the problem can be solved in a rather straightforward way
using the traditional radiative transfer techniques
(Rowan-Robinson~1980; Yorke~1980; Wolfire \& Cassinelli~1986; Rogers
\& Martin~1986; Ivezi\'c~\& Elitzur~1997). However, for more general
two- and three-dimensional geometries the problem is much more
difficult, and obtaining an exact solution becomes much harder with
these conventional techniques (Efstathiou \& Rowan-Robinson~1990,
1991; Sonnhalter et al.~1995; Men'shchikov \& Henning~1997; Steinacker
et al.~2003). This can be overcome with the Monte Carlo method, where
there are no geometrical restrictions.  The Monte Carlo method for
radiative equilibrium radiative transfer calculations was pioneered
more than two decades ago (Lefevre et al.~1982, 1983). Since then,
various new techniques, optimizations and extensions have been
proposed (Lucy~1999; Bjorkman
\& Wood~2001; Misselt et al.~2001; Ercolano et al.~2003b; Niccolini,
Woitke \& Lopez~2003) and the method has been applied to widely
different environments, including stellar atmospheres (Lucy~1999; Wood
et al.~2002), dusty galaxies (Bianchi et al.~2000a; Misselt et
al.~2001), planetary nebulae (Ercolano et al.~2003a) and protostellar
cores (Wolf et al.~1999; Stamatellos \& Whitworth~2003).

One of these optimizations, proposed by Bjorkman \& Wood~(2001,
hereafter BW), seems particularly interesting. These authors describe
a technique in which iteration, which is an undesirable but necessary
ingredient of the standard radiative equilibrium Monte Carlo
techniques, can be avoided. However, it is not completely clear, in
our opinion, whether the frequency distribution adjustment (hereafter
FDA) technique, which the BW method employs, is an exact rigorous
method, or whether it is just a good and convenient
approximation. This narrow distinction is quite important, not only
from an academic point of view. For example, Ercolano et al.~(2003b)
have developed an iterative Monte Carlo code for radiative equilibrium
calculations in which the opacity of the obscuring medium depends on
its temperature. They adopt the FDA technique in each iteration
step. Any small deviations from the correct answer could culminate in
larger errors after performing the same method at each iteration step.

It is therefore important to obtain a thorough theoretical
understanding of the FDA method, and to investigate its advantages and
limitations. This is the goal of the current paper. In section~2, we
will describe the problem of radiative and thermal equilibrium Monte
Carlo radiative transfer, and we focus on the FDA procedure proposed
to avoid iteration. In section~3, we investigate the basis of the FDA
procedure. We do this in a broader framework, and investigate the
general problem of how a sampled distribution can be adjusted to a
desired distribution by adding data points sampled from an adjustment
distribution. We apply this to radiative equilibrium Monte Carlo
radiative transfer in section~4, and compare the results with the
method of BW. In section~5 we discuss the results and present the
advantages and limitations of the FDA procedure.

\section{Radiative equilibrium Monte Carlo radiative transfer}

The basic idea of Monte Carlo radiative transfer is that a very large
number of $L$-packets (often called photons or photon packets) are
followed individually throughout the system while being emitted,
scattered or absorbed. Each step in the lifetime of a single
$L$-packet is governed by random events. First, we divide the total
luminosity $L$ emitted by the radiation sources into a very large
number $N$ of packets of equal luminosity $\delta L=L/N$. Next, the
dust medium is divided into a number of dust cells, each of them
representing a physical dust entity with a specific mass $M$ and
temperature $T$ (the latter still has to be determined). We then start
the actual Monte Carlo simulation by launching each of these $N$
packets randomly in the dusty medium.  Each of them is assigned a
random initial position $\bfr$ according to the geometry of the
sources, a random initial propagation direction $\bfk$ and a random
frequency $\nu$. During its journey through the dusty medium, the
position and propagation direction of each $L$-packet changes, until
it either leaves the system (and can be detected) or is absorbed by a
dust grain.

\subsection{Solution using iteration}

In the standard application of the radiative equilibrium Monte Carlo
calculations, we launch all direct source $L$-packets, follow them
until they leave the system or are absorbed, and record the cell in
which each absorption takes place.  After launching and following all
source $L$-packets, we calculate the dust temperature of each cell
from the cell mass and the amount of absorbed luminosity. If $k$
packages have been absorbed in a dust cell, the temperature of this
cell can be determined from the requirement that the total absorption
rate ($\Gamma_{\text{abs}}$) in the cell must equal the total emission
rate ($\Gamma_{\text{em}}$) from the cell. The former can be estimated
by multiplying the number of $L$-packets that have been absorbed with
the luminosity fraction $\delta L$ per $L$-packet, i.e.
\begin{equation}
    \Gamma_{\text{abs}}
    =
    k\,\delta L
    =
    k\,\frac{L}{N},
    \label{absorptionrate}
\end{equation}
whereas the latter is obtained from
\begin{equation}
    \Gamma_{\text{em}}
    =
    4\pi\, M \int_0^\infty
    \kappa_\nu B_\nu(T)\,\txd\nu.
\end{equation}
From the requirement that the absorption and emission rate balance
each other,
\begin{equation}
    \Gamma_{\text{em}}
    =
    \Gamma_{\text{abs}},
    \label{radeq}
\end{equation}
the new temperature $T$ of the cell can be determined. In order to
conserve radiative equilibrium, we must launch a new set of $k$
$L$-packets, which should be distributed according to the emissivity
of the dust cell,
\begin{equation}
    j_\nu = \kappa_\nu B_\nu(T).
\end{equation}
Hence each re-emitted $L$-packet must be assigned a frequency
determined from the probability density function (hereafter PDF)
\begin{equation}
    p(\nu)\txd\nu
    =
    \frac{\kappa_\nu B_\nu(T)\,\txd\nu}{\int_0^\infty
    \kappa_\nu B_\nu(T)\,\txd\nu}.
\end{equation}
By emitting these $k$ $L$-packets we assure that radiative equilibrium
is satisfied in each cell.

Each of these re-emitted $L$-packets is again followed through the
dust grid until either it is absorbed or it leaves the system.  Again,
we record the number of absorption events in each cell.  This allows
us to update the temperature of each cell, to determine the new
emissivity, etc. This procedure naturally leads to an iteration, which
stops when the temperature of each dust cell converges.

Unfortunately, Monte Carlo radiative transfer is a numerically
demanding method, and so it would be very useful if somehow the
iteration in this method could be either accelerated or avoided
completely. Lucy~(1999) has proposed a device to optimize the
iteration. He argues that a much faster convergence can be achieved by
applying a temperature correction and re-emission immediately after
each individual absorption event. Thus, every single time an
$L$-packet is absorbed, the temperature $T$ of the cell is updated to
a new temperature $T+\Delta T$ and a new $L$-packet is re-emitted with
the new emissivity of the cell $\kappa_\nu B_\nu(T+\Delta T)$. This
$L$-packet is then followed until either it leaves the system or it is
absorbed again. Hence the re-emitted $L$-packets enter the radiative
transfer alongside the source $L$-packets, and also contribute to the
heating of the dust. As a consequence, the temperature distribution
and the radiation field of the system after the last $L$-packet leaves
the system are closer to the equilibrium state, and fewer iterations
are necessary to reach convergence.

\subsection{Solution using the FDA method}
\label{bw.sec}

Even with the improvements made by Lucy~(1999), iteration remains
necessary. Indeed, during the calculation, the temperature in each
cell gradually increases, and the $L$-packets re-emitted in the
beginning of the simulation are hence emitted from an incorrect
frequency distribution (corresponding to a temperature which is too
low). An extension of the idea of Lucy~(1999) has been proposed by
BW. Their aim is to correct, during each re-emission event, for the
incorrect frequency distribution of the $L$-packets that have been
re-emitted before. If this succeeds, then the system is in radiative
and thermal equilibrium at all moments during the simulation, and
after the last $L$-packet leaves the system, the correct temperature
distribution and spectrum are obtained without any iteration at all.

Consider some stage in the Monte Carlo process, when $k$ $L$-packets
have already been absorbed and re-emitted in a certain cell. The
temperature of the cell has been gradually increasing from zero to
$T$, and assume we somehow have managed to fine-tune the re-emission
frequencies such that they {\em all} correspond to the emissivity
$\kappa_\nu B_\nu(T)$. Now assume a $(k+1)$'th $L$-packet is absorbed,
which increases the dust cell temperature to $T+\Delta T$, so that the
emissivity changes to $\kappa_\nu B_\nu(T+\Delta T)$. To preserve
radiative equilibrium, a new $L$-packet must be emitted with
luminosity $\delta L$. Its frequency must be chosen in such a way that
the ensemble of the $(k+1)$ re-emitted $L$-packets from this cell
correspond to the new emissivity $\kappa_\nu B_\nu(T+\Delta T)$. BW
argue that this can be satisfied if the $(k+1)$'th frequency
corresponds to the difference emissivity
\begin{equation}
    \Delta j_\nu
    =
    \kappa_\nu \Bigl[ B_\nu(T+\Delta T)-B_\nu(T) \Bigr].
\end{equation}
Hence the $(k+1)$'th frequency should hence be drawn from the
normalized PDF
\begin{equation}
    \tilde{q}(\nu)\txd\nu
    =
    \frac{\kappa_\nu [B_\nu(T+\Delta T)-B_\nu(T)] \txd\nu}
    {\int_0^\infty \kappa_\nu [B_\nu(T+\Delta T)-B_\nu(T)] \txd\nu}.
\label{qBW}
\end{equation}
Note that this PDF is positive everywhere, because the Planck function
is a monotonically increasing function of temperature. When the
difference $\Delta T$ between the two temperatures is small, this
expression can be approximated by
\begin{equation}
    \tilde{q}(\nu)\txd\nu
    \approx
    \frac{\kappa_\nu B'_\nu(T)\,\txd\nu}
    {\int_0^\infty\kappa_\nu B'_\nu(T)\,\txd\nu},
\end{equation}
where $B'_\nu(T)$ is the temperature derivative of the Planck
function.

BW have implemented their technique and tested it on the set of
benchmark models presented by Ivezi\'{c} et al.~(1997). They find good
agreement with the benchmark results, and conclude that their FDA
method works well. Furthermore, we have tested this method against the
thermodynamic equilibrium test and shown that the method is indeed
accurate: the temperature structure of a system embedded in an
isotropic black-body radiation field was found to be uniform, with an
accuracy which could be reduced to less than 0.1~K by simulating
enough $L$-packets (Stamatellos \& Whitworth~2003).

Two arguments, however, made us doubt whether this method is really an
exact rigorous method, or whether it is just a very good and
convenient approximation:
\begin{enumerate}
\item Equation~(\ref{qBW}) suggests that one can always adjust the
spectrum of $k$ emitted frequencies from one PDF (corresponding to
$T$) to another (corresponding to $T+\Delta T$) by simply adding one
single frequency. Intuitively we expect that this is possible if the
difference between these two temperatures is small, but that it will
be very hard or impossible if the temperature difference is large. For
example, when the temperature is very low, most of the radiation will
be emitted at far-infrared or sub-millimeter wavelengths. When the
temperature increase is sufficiently high, the new emissivity requires
that most $L$-packets are emitted at optical or near-infrared
wavelengths. Consequently, there is a limit on the temperature
increase allowed by this method. \item It seems logical that adjusting
a sampled distribution by adding one more data point is much harder if
many frequencies have been sampled before, than when only a small
number of frequencies have been sampled. Therefore, one would in
general expect that the ''adjustment PDF'' $\tilde{q}(\nu)\txd\nu$
would be an explicit function of the number $k$ of previously emitted
$L$-packets.
\end{enumerate}
In order to investigate this in detail, we study the issue of
adjusting the behavior of a set of data points by adding new data
points in a more general context in the next section, and apply the
results to radiative equilibrium Monte Carlo radiative transfer
afterwards.

\section{Adjusting a previously sampled distribution}
\label{adjust.sec}

Assume $p(x)\txd x$ and $r(x)\txd x$ are two arbitrary PDFs, and
consider the following question: is it possible to construct a third
PDF $q(x)\txd x$ which satisfies the condition that any set of $n+m$
data points, where the first $n$ data points are drawn randomly from
$p(x)\txd x$ and the last $m$ ones are drawn randomly from $q(x)\txd
x$, represents a set of data drawn randomly from $r(x)\txd x$~? In
other words, can we find an ''adjustment PDF'' that adjusts an
arbitrary original PDF into an arbitrary altered PDF~?

The solution to this problem is based on the definition that the
probability that any data point lies between $x$ and $x+\txd x$, when
drawn randomly from a PDF $p(x)\txd x$, equals $p(x)\txd x$.  Hence,
when we draw $n$ random data points from this PDF, the expected number
of them between $x$ and $x+\txd x$ is $n\,p(x)\txd x$. Applying this
to the total data set of $m+n$ data points, we can find the form of
the required PDF. As we can simply add the number of data points in
each interval, the expected number of data points between $x$ and
$x+\txd x$ is given by
\begin{equation}
    \txd N(x)
    =
    n\,p(x)\txd x
    +
    m\,q(x)\txd x.
\label{dN}
\end{equation}
On the other hand, this sample of $m+n$ data points should
represent a sample drawn randomly from the PDF $r(x)\txd x$, and
therefore we can also write
\begin{equation}
    \txd N(x)
    =
    (n+m)\,r(x)\txd x.
\end{equation}
From these two expressions, we find that the adjustment PDF should
have the form
\begin{equation}
    q(x)\txd x
    =
    \left(1+\frac{n}{m}\right) r(x)\txd x
    -
    \frac{n}{m}\,p(x)\txd x.
\label{qdef}
\end{equation}
However, it is not always guaranteed that this function is a valid
PDF, because there is no {\em a priori} guarantee that it is positive
over its entire domain. If this function does become negative, this
means that it is impossible to adjust the PDF $p(x)\txd x$ to
$r(x)\txd x$ by adding $m$ data points.

Whether or not it is possible to adjust one PDF into another depends
both on the particular PDFs involved and on the ratio $k
\equiv n/m$ of the number of data points already sampled to the
number of data points which can be added. Logically, it will be easier
to transform one PDF to another if the distributions are very similar,
than when they are very different. Also, it will in general be easier
to change one distribution into another one if many data points can be
added to a few data points sampled before ($m\gg n$), than when only a
few new data points can be added to a large set of already existing
data points ($n\gg m$).

In appendix~A, we demonstrate the adjustment of one PDF to another for
gaussian distributions. If we want to adjust the original gaussian
distribution into another gaussian distribution with the same
dispersion but a different mean, we show that the resulting adjustment
PDF will always be negative at some point, no matter how small the
shift in the mean or the relative number of data points to be
added. On the other hand, if we want to alter the original gaussian
PDF to a gaussian PDF with the same mean but a larger dispersion, we
find that this is possible in some cases. When the relative number of
data points to add is too small or when the dispersion increase is too
large, however, the adjustment is not possible.

\begin{figure*}
\centering
\includegraphics[clip,bb=315 29 496 613,angle=-90,width=\textwidth]{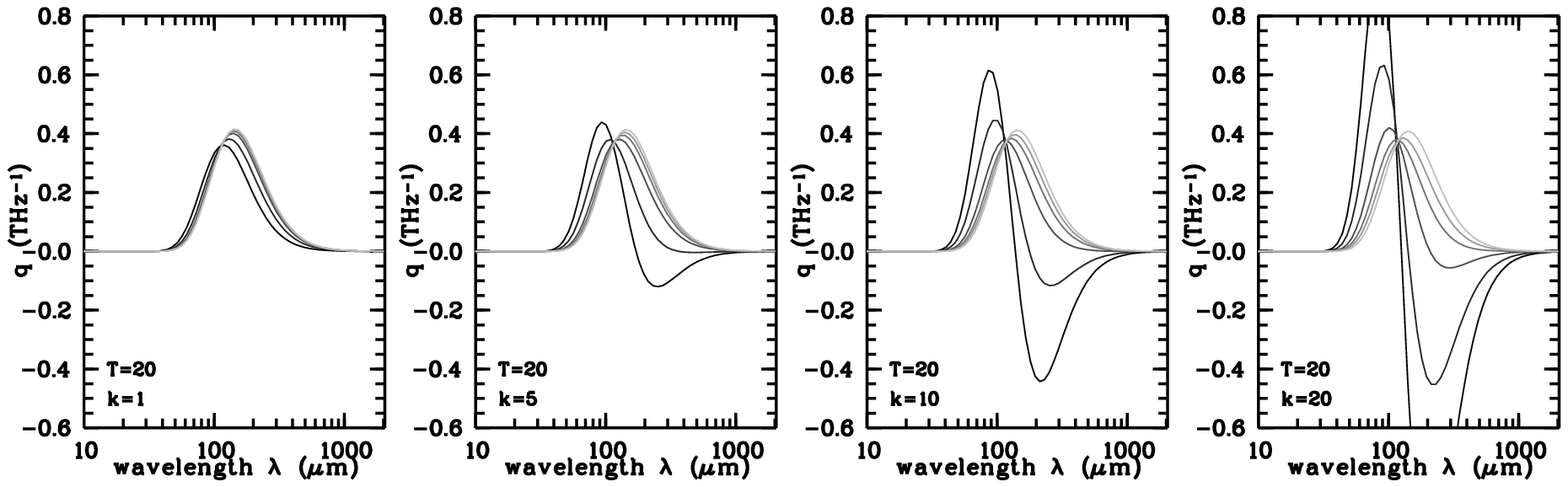}
\caption{Examples of the adjustment PDF (\ref{qwe}), corresponding
to an opacity function $\kappa_\nu\propto\nu^2$ and $T=20$~K. The four
different panels show the adjustment PDF for various values of $k$. In
each panel, the different curves correspond to different values of the
temperature increase, corresponding to $\Delta T=0.02$~K (light grey),
0.1~K, 0.2~K, 0.4~K, 1~K and 2~K (black). For small values of $k$ and
$\Delta T$, the adjustment PDF is a positive function, and adjustment
is possible. For large values of $k$ and/or $\Delta T$, the adjustment
PDF is negative at long wavelengths, and adjustment is not possible.}
\label{plotq.eps}
\end{figure*}

\section{Application to radiative equilibrium Monte Carlo radiative transfer}

\subsection{Determination of the adjustment PDF}

With the general theorem from the previous section, we can return to
the radiative equilibrium problem addressed in
Section~{\ref{bw.sec}}. What we need to do is to adjust a sample of
$k$ frequencies drawn randomly from the PDF
\begin{equation}
    p(\nu)\txd\nu
    =
    \frac{\kappa_\nu B_\nu(T)\,\txd\nu}{\int_0^\infty
    \kappa_\nu B_\nu(T)\,\txd\nu}
\end{equation}
to a sample of $k+1$ frequencies drawn from the PDF
\begin{equation}
    r(\nu)\txd\nu
    =
    \frac{\kappa_\nu B_\nu(T+\Delta T)\,\txd\nu}{\int_0^\infty
    \kappa_\nu B_\nu(T+\Delta T)\,\txd\nu},
\end{equation}
by just adding one single frequency $L$-packet. Applying the general
formula~(\ref{qdef}), we find that the last emitted $L$-packet should
have a frequency drawn from the adjustment PDF
\begin{equation}
    q(\nu)\txd \nu
    =
    (1+k)\,\frac{\kappa_\nu B_\nu(T+\Delta T)\,\txd\nu}
    {\int_0^\infty\kappa_\nu B_\nu(T+\Delta T)\,\txd\nu}
    -
    k\,\frac{\kappa_\nu B_\nu(T)\,\txd\nu}
    {\int_0^\infty \kappa_\nu B_\nu(T)\,\txd\nu}.
    \qquad\quad
\label{qwe}
\end{equation}
This formula can be easily interpreted in a physical way. When the
$(k+1)$'th $L$-packet is absorbed, the temperature rises from $T$ to
$T+\Delta T$. This last $L$-packet should then be re-emitted with a
frequency sampled from the PDF corresponding to the new temperature,
i.e.
\begin{equation}
    P_{k+1}(\nu)\txd \nu
    =
    \frac{\kappa_\nu B_\nu(T+\Delta T)\,\txd\nu}
    {\int_0^\infty\kappa_\nu B_\nu(T+\Delta T)\,\txd\nu}.
\end{equation}
Additionally, we should account for the $L$-packets that were emitted
previously with the incorrect frequency. Thus, we should re-emit the
previous $k$ $L$-packets with a frequency sampled from the difference
between the new and the old PDF,
\begin{equation}
    P_j(\nu)\txd\nu
    =
    \frac{\kappa_\nu B_\nu(T+\Delta T)\,\txd\nu}
    {\int_0^\infty\kappa_\nu B_\nu(T+\Delta T)\,\txd\nu}
    -
    \frac{\kappa_\nu B_\nu(T)\,\txd\nu}
    {\int_0^\infty\kappa_\nu B_\nu(T)\,\txd\nu}.
    \qquad
    (j=1\ldots k)
\end{equation}
The above procedure is equivalent to emitting just one $L$-packet with
the combined PDF as given by equation~(\ref{qwe}).

However, it is not guaranteed that this function corresponds to a
valid PDF, because it can be negative. This is illustrated in
figure~{\ref{plotq.eps}}, where we plot the adjustment PDF $q(\nu)$
for $T=20$~K and different values of the parameters $\Delta T$ and
$k$. The opacity function adopted is a simple power law of frequency,
$\kappa_\nu \propto \nu^2$. For small values of $k$, the adjustment
PDF is positive for a large range of temperature shifts. However, when
the number of previously generated frequencies $k$ is high, the
adjustment PDF will be positive only for the smallest values of
$\Delta T$. This behavior is also illustrated in
figure~{\ref{modplanckq.eps}}, where we explicitly show the region in
$(k,T,\Delta T)$ parameter space where adjustment is possible. The
solid diagonal lines (different lines are shown for different values
of $T$, but there is hardly any dependence on $T$) mark the border
between positive and negative adjustment PDFs. For each number of data
points $k$ and each temperature $T$, there is a maximum temperature
increase $\Delta T_{\text{max}}(T,k)$ that can be allowed for adding
one $(k+1)$'th frequency $L$-packet. As expected, this maximum
temperature increase strongly decreases with increasing $k$.

\subsection{Comparison with the BW adjustment PDF}
\label{compBW.sec}

The calculation in the previous subsection showed that the adjustment
PDF (\ref{qBW}) proposed by BW is in general not the same as the
adjustment PDF~(\ref{qwe}). Moreover, we have shown that the latter
does not represent a proper PDF for all of the parameters $(k,T,\Delta
T)$. Indeed, for a given $k$ and $T$ there is a critical value of
$\Delta T$, above which the adjustment PDF becomes negative. This
observation suggests that the method by BW is an approximation rather
than a rigorous solution, and even that the exact FDA procedure in
Monte Carlo radiative transfer might not always be possible.

\begin{figure}
\centering
\includegraphics[clip,bb=184 492 414 687,width=0.45\textwidth]{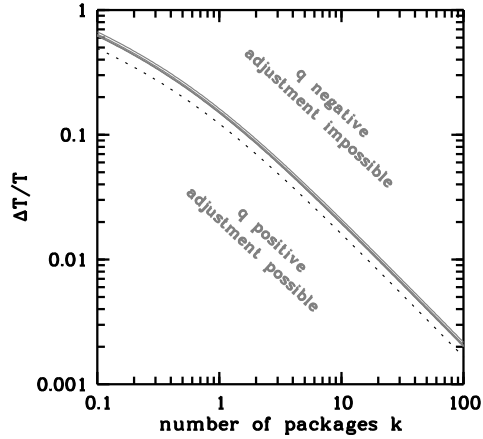}
\caption{Region in $(k,T,\Delta T)$ parameter space in which the
adjustment PDF~(\ref{qwe}) is positive. The opacity function is again
$\kappa_\nu\propto\nu^2$. Different solid grey lines correspond to
different values of the temperature $T$, but there is hardly any
dependence on $T$. Each line shows for a given $T$ and $k$ the maximum
temperature increase $\Delta T$ such that the adjustment PDF is
positive. Hence, underneath the solid grey lines, adjustment is
possible, above these lines, adjustment is impossible. The dotted line
indicates the relative temperature increase expected from the
radiative equilibrium requirement (see section~{\ref{compBW.sec}}).}
\label{modplanckq.eps}
\end{figure}

It is important to realize however, that not the entire $(k,T,\Delta
T)$ parameter space will be covered during a simulation. Indeed, if a
dust cell has a given temperature $T$ after absorbing and re-emitting
$k$ $L$-packets, and the $(k+1)$'th $L$-packet is absorbed, the
temperature rise $\Delta T$ is determined by the requirement of
radiative equilibrium~(\ref{radeq}),
\begin{equation}
    \int_0^\infty
    \kappa_\nu B_\nu(T+\Delta T)\,\txd\nu
    =
    \frac{(k+1)\,\delta L}{4\pi M}.
\label{radeqexpl}
\end{equation}
For the opacity function $\kappa_\nu \propto \nu^2$, the
temperature increase $\Delta T$ can be calculated explicitly for
each $T$ and $k$. Using the expression
\begin{equation}
    \int_0^\infty \nu^2 B_\nu(T)\,\txd\nu \propto T^6,
\end{equation}
one obtains after some algebra
\begin{equation}
    \Delta T(T,k)
    =
    T\left[\left(1+\frac{1}{k}\right)^{1/6}-1\right].
\end{equation}
This function is plotted as the dotted line in
figure~{\ref{modplanckq.eps}}. This plot shows that $\Delta
T(T,k)<\Delta T_{\text{max}}(T,k)$, i.e. at every step in the Monte
Carlo simulation, the temperature increase is always small enough such
that the adjustment PDF is positive. This proves that the FDA method
in radiative equilibrium Monte Carlo radiative transfer works.

Knowing that the adjustment PDF (\ref{qwe}) will always be
positive for the expected temperature increase, we can compare the
results with those of BW. From equation~(\ref{radeqexpl}), we
obtain that
\begin{equation}
    \frac{\int_0^\infty\kappa_\nu B_\nu(T+\Delta T)\,\txd\nu}
    {\int_0^\infty\kappa_\nu B_\nu(T)\,\txd\nu}
    =
    \frac{k+1}{k}.
\end{equation}
If we use this expression to eliminate $k$ from the adjustment
PDF~(\ref{qwe}), we obtain
\begin{equation}
    q(\nu)\txd\nu
    =
    \frac{\kappa_\nu [B_\nu(T+\Delta T)-B_\nu(T)] \txd\nu}
    {\int_0^\infty \kappa_\nu [B_\nu(T+\Delta T)-B_\nu(T)] \txd\nu}
    \equiv
    \tilde{q}(\nu)\txd\nu,
\end{equation}
i.e.\ we recover the adjustment PDF proposed by BW.

\section{Discussion}

\subsection{Efficiency of the method}

The analysis in the previous section demonstrates that the FDA
procedure proposed by BW provides the correct distribution of
frequencies of the re-emitted $L$-packets. This demonstrates that the
FDA algorithm is robust, and that at every step in the Monte Carlo
simulation, the spectrum of the re-emitted $L$-packets is in agreement
with the temperatures of the dust medium. As it avoids computationally
costly iteration, the method is probably the most efficient way of
doing thermal equilibrium radiative transfer calculation for arbitrary
geometries.

One competitor might be the method pioneered by Niccolini et
al.~(2003), in which the determination of the temperature structure is
separated from the Monte Carlo radiative transfer procedure. With the
results of a single radiative transfer run with an initially guessed
temperature distribution, they determine the correct temperature
distribution and then scale the radiation field by the corresponding
factors. Although their method also avoids performing several Monte
Carlo simulation runs in an iterative loop, it still contains an
iterative procedure to determine the temperature distribution. When
the number of cells is small, this iteration is fairly quick, and this
method is probably competitive with the FDA method of BW. When the
number of cells is large, however, which may be necessary in realistic
three-dimensional geometries without obvious symmetries, this
iterative loop to determine the temperature distribution might be very
difficult and time-consuming. This probably makes the FDA method the
most efficient method for the general three-dimensional radiative
equilibrium radiative transfer problem.

\subsection{Extensions and optimizations}

Although we have demonstrated that the method presented by BW
represents the correct way to adjust the frequency distribution of the
previously emitted $L$-packets, it should be kept in mind that the
correct behavior is due to the additional requirement of local thermal
and radiative equilibrium, as given in
equation~(\ref{radeqexpl}). Whenever deviations are made from this
requirement, the extended technique might fail.

\subsubsection{Photon weighting}
\label{weighted.sec}

We have assumed so far that all $L$-packets are luminosity packets
with the same luminosity $\delta L$. When such an $L$-packet is
absorbed, a new $L$-packet with a different frequency but with the
same luminosity must be re-emitted. As such, the Monte Carlo
simulation can be very inefficient in regions with low density. One
option is to make the cells bigger such that more absorptions occur,
but this decreases the spatial resolution. A more intelligent option
is to drop the requirement that all $L$-packets must have the same
luminosity, or equivalently, to introduce $L$-packet weights. By doing
this, devices which reduce Poisson noise can be built into the Monte
Carlo code, such as the principle of forced interaction or a peel-off
technique (e.g.~Cashwell \& Everett~1959; Witt~1977; Yusef-Zadeh,
Morris \& White~1980; Niccolini et al.~2003).

We should consider whether such techniques are compatible with the FDA
technique. To answer this question, assume that $k$ $L$-packets with
luminosity $\delta L$ have been absorbed in a cell, that the
temperature of the cell is $T$, and that a $k+1$'th $L$-packet is
absorbed with a different luminosity, $w\,\delta L$.  Clearly, an
$L$-packet with this luminosity will have to be re-emitted, but what
will be the shape of the adjustment PDF in this case~? In
section~{\ref{adjust.sec}} we saw that the adjustment PDF
corresponding to adding $m$ data points to a set of $n$ existing data
points only depends on the {\em relative} number $k=m/n$ of existing
data points. Translating this to the current case means that the
adjustment PDF corresponding to adding an $L$-packet with luminosity
$w\,\delta L$ to a set of $k$ $L$-packets with luminosity $\delta L$,
is equivalent to adding one $L$-packet with luminosity $\delta L$ to a
set of $k/w$ $L$-packets with luminosity $\delta L$. The same
formula~(\ref{qwe}) will be appropriate but with $k$ replaced by
$k/w$. We should in general not consider $k$ as an integer number
corresponding to the number of $L$-packets absorbed by the dust cell
before the last absorption event, but rather as a real number
corresponding to the ratio of the luminosity of the last absorbed
$L$-packet to the total luminosity absorbed by the dust cell before
the last absorption event.  Because $k$ has naturally the same meaning
in equation~(\ref{absorptionrate}) for the absorption rate, the BW
adjustment PDF will still be valid. $L$-packet weighting can therefore
be combined easily with the FDA method.

\subsubsection{Additional heating sources}

The technique described in this paper considers the calculation of the
temperature distribution of dust grains, which are in radiative
equilibrium with the radiation field. In realistic astrophysical
situations, heating by the ambient radiation field is often not the
only source of dust heating. Dust grains can be heated by a variety of
other astrophysical processes, such as viscous or compressional
heating, collisions with hot electrons in an X-ray halo or collisions
with cosmic rays. In this case, the condition of radiative
equilibrium~(\ref{radeq}) is not satisfied, and should now be replaced
by a more general equation
\begin{equation}
    \Gamma_{\text{em}}
    =
    \Gamma_{\text{abs}} + \Gamma_{\text{add}},
\end{equation}
where $\Gamma_{\text{add}}$ represents the amount of energy per unit
time (the luminosity) gained by the dust cell due to other
processes. If we assume that this factor in general depends on the
position, size, etc.\ of the dust cell, but not on its temperature, it
remains constant during the Monte Carlo simulation.

Although radiative equilibrium is not satisfied, it is straightforward
to see that the FDA method is fully compatible with such additional
heating sources. Indeed, it suffices, as in
section~{\ref{weighted.sec}, to consider $k$ as the ratio of the
luminosity of the last absorbed $L$-packet to the total luminosity
gained by the dust cell before the last absorption event. In
particular, this means that the temperature $T_0$ of the dust cell at
the beginning of the simulation is not zero, but is determined by
\begin{equation}
    \int_0^\infty \kappa_\nu B_\nu(T_0)\,\txd\nu
    =
    \frac{\Gamma_{\text{add}}}{4\pi\,M}.
\end{equation}
The FDA method can therefore be combined easily with additional
heating sources.

\subsubsection{Very small dust grains}

Another case where condition~(\ref{radeqexpl}) is not satisfied occurs
when the contribution of small dust grains is important. These dust
grains undergo transient heating to temperatures well beyond the
equilibrium temperature (Guhathakurta
\& Draine~1989). This means that the grains within a dust cell
have a range of time-dependent temperatures characterized by a
temperature probability function $P(T)$ rather than a single
equilibrium temperature. Although more complex physics are involved
here, their inclusion in Monte Carlo radiative transfer calculation is
in principle similar to the normal radiative equilibrium radiative
transfer (e.g.\ Misselt et al.~2001). So in principle, the FDA
technique could also be applied to this more difficult
problem. However, it can be expected that the relative temperature
increase of small dust grains may be so large that negative adjustment
PDFs are obtained. Therefore, in the case of transient heating by
small dust grains, iteration can probably not be avoided.

\subsubsection{An alternative estimate for the absorption rate}

In this Monte Carlo method, we have estimated the absorption rate by
multiplying the number of $L$-packets that have been absorbed with the
luminosity fraction $\delta L$ per $L$-packet, as in
equation~(\ref{absorptionrate}). However, this way of estimating the
absorption rate performs rather poorly in low density environments. A
better way of estimating the absorption rate is to use its direct link
to the mean intensity of the radiation field,
\begin{equation}
    \Gamma_{\text{abs}}
    =
    4\pi\,M
    \int_0^\infty \kappa_\nu J_\nu \txd\nu.
\end{equation}
As Lucy~(1999) argues, the mean intensity in a given cell can be
estimated through its relation to the energy density of the radiation
field. We only have to add a path length counter in each cell and
determine the total path length covered by all $L$-packets in the cell
(see also Niccolini et al.~2003). The advantage of this method is that
all $L$-packets entering the dust cell will contribute to the estimate
of the absorption rate. Since generally only a small number of the
$L$-packets entering a cell will be absorbed, it is clear that this
method will give a better estimate of the absorption rate, and hence
of the temperature of the dust cells.

If we estimate the absorption rate in this way however, the relation
(\ref{radeqexpl}) will not generally be valid anymore, and there is no
reason why the adjustment PDF~(\ref{qwe}) should be equal to the BW
PDF~(\ref{qBW}). Therefore, for the FDA method to work correctly, it
is necessary to estimate the absorption rate as in
equation~(\ref{absorptionrate}), which might be rather inefficient in
environments with a low density. When performing radiative equilibrium
Monte Carlo simulations, it is hence strongly recommended to construct
the dust cells in such a way that the absorption probability is more
or less equal in each of the cells.

\section{Conclusions}

We have evaluated critically the frequency distribution adjustment
(FDA) method, a technique proposed by Bjorkman \& Wood~(2001) to
optimize radiative equilibrium Monte Carlo radiative transfer
simulations.

We first investigated the more general problem of trying to adjust
the spectrum of a set of data points sampled from an arbitrary
distribution by adding extra data points, such that the combined
data set appears to be sampled from an arbitrary third
distribution. We have determined the general shape this
distribution must have, and shown that it is not always possible
to adjust a set of data points in an arbitrary way. Whether this
is possible, depends on the shape of the distributions and on the
relative number of data points that have already been sampled.

We use this general theorem to investigate the theoretical basis of
the FDA method. We demonstrate that the FDA method provides the
correct frequency distribution for the re-emitted $L$-packet, because
of the additional requirement of radiative equilibrium. We also show
that the method can be easily extended with the use of weighted
$L$-packets, and that additional heating mechanisms can be included in
the simulation without violating the FDA method. However, the method
may fail if small dust grains are included, or if the absorption rate
is estimated in an alternative way.

\appendix
\section{Example of adjusting a gaussian distribution}

Suppose that we have sampled a set of $n$ data points from a
gaussian PDF,
\begin{equation}
    p(x)\txd x
    =
    \frac{1}{\sqrt{2\pi}\sigma}\,
    \exp\left[-\frac{(x-\mu)^2}{2\sigma^2}\right]
    \txd x,
\end{equation}
and we want to add a number $m=n/k$ data points such that the
total data set represents a sample taken from a different gaussian
PDF.

First, assume the required PDF is a gaussian with the same
dispersion, but we shift the mean over a distance $\Delta\mu>0$
\begin{equation}
    r(x)\txd x
    =
    \frac{1}{\sqrt{2\pi}\sigma}\,
    \exp\left[-\frac{(x-\mu-\Delta\mu)^2}{2\sigma^2}\right]
    \txd x.
\end{equation}
Using expression~(\ref{qdef}), we find the resulting adjustment
PDF $q(x)\txd x$. It is a straightforward exercise to see that
this function will always be negative in the range
\begin{equation}
    x
    <
    \mu +
    \frac{\sigma^2}{\Delta\mu}\ln\left(\frac{k}{1+k}\right)
    -
    \frac{\Delta\mu}{2}.
\end{equation}
So, no matter how small $k$ or $\Delta\mu$, the function $q(x)\txd
x$ will never represent a proper PDF, such that it is (in theory)
impossible to shift the mean of a gaussian distribution by adding
more data points.

Next, assume the required PDF is a gaussian with the same mean,
but with a different dispersion $\sigma'=(1+\delta)\sigma$,
\begin{equation}
    r(x)\txd x
    =
    \frac{1}{\sqrt{2\pi}(1+\delta)\sigma}\,
    \exp\left[-\frac{(x-\mu)^2}{2(1+\delta)^2\sigma^2}\right]
    \txd x.
\end{equation}
Again, the appropriate adjustment PDF $q(x)\txd x$ can be obtained
using equation~(\ref{qdef}). A similar exercise shows that this
function becomes negative if,
\begin{equation}
    (x-\mu)^2
    <
    \frac{2(1+\delta)^2\sigma^2}{(1+\delta)^2-1}
    \ln\left[\frac{(1+\delta)k}{1+k}\right].
\end{equation}
This condition will, however, never be satisfied if the right-hand
side of this equation is negative, i.e.\ when $\delta<1/k$. In
this case, the function $q(x)$ will therefore be positive for all
values of $x$ and represents a valid PDF.

\end{document}